\documentclass[aps,prl,reprint,superscriptaddress]{revtex4-2}

\usepackage{amsmath,amsthm,amssymb}
\usepackage{dcolumn,bm,hyperref}
\usepackage{graphicx,subfigure,verbatim}
\usepackage{txfonts}    
\usepackage{xcolor}

\usepackage{pdfpages}
\makeatletter\AtBeginDocument{\let\LS@rot\@undefined}\makeatother

\renewcommand{\vec}[1]{{\bm{\mathrm{#1}}}}

\DeclareMathOperator{\Tr}{Tr}

\newcommand{\bra}[1]{\langle{#1}|}
\newcommand{\ket}[1]{|{#1}\rangle}

\usepackage{xcolor}

\begin{document}
	\title{Orbital Dynamics in Centrosymmetric Systems}
	
	\author{Seungyun Han}
	\affiliation{Department of Physics, Pohang University of Science and Technology, Pohang 37673,Korea}
	
	\author{Hyun-Woo Lee}
	\email{hwl@postech.ac.kr}
	\affiliation{Department of Physics, Pohang University of Science and Technology, Pohang 37673,Korea}
	
	\author{Kyoung-Whan Kim}
	\email{kwk@kist.re.kr} 
	\affiliation{Center for Spintronics, Korea Institute of Science and Technology , Seoul 02792, Korea}
	
	\pacs{\quad}
	
	\date{\today}
	
	\begin{abstract}
		Orbital dynamics in time-reversal-symmetric centrosymmetric systems is examined theoretically. Contrary to common belief, we demonstrate that many aspects of orbital dynamics are qualitatively different from spin dynamics because the algebraic properties of the orbital and spin angular momentum operators are different. This difference generates interesting orbital responses, which do not have spin counterparts. For instance, the orbital angular momentum expectation values may oscillate even without breaking neither the time-reversal nor the inversion symmetry. Our quantum Boltzmann approach reproduces the previous result on the orbital Hall effect and reveals additional orbital dynamics phenomena, whose detection schemes are discussed briefly. Our work will be useful for the experimental differentiation of the orbital dynamics from the spin dynamics.
	\end{abstract}
	
	\maketitle
	
	{\it Introduction.--}  Orbital angular momentum (OAM) of electrons is a fundamental degree of freedom in condensed matter systems. Not only is it crucial for equilibrium properties of solids, but also it may generate various nonequilibrium phenomena, such as the orbital Hall effect~\cite{bernevig2005b,tanaka2008,kontani2008}, the orbital Edelstein effect~\cite{Yoda2015,Chen2018,salemi2019,ding2020,ding2021}, and the orbital torque~\cite{go2020,zheng2020,tazaki2020,kim2021}. Its coupling with other degrees of freedom like spin and valley~\cite{schaibley2016} results in the strong spin Hall effect in transition metals~\cite{kontani2009,go2018}, light-valley interaction~\cite{mak2018}, and magnetoresistance~\cite{zhou2019,ko2020}. Recently, the generality~\cite{go2018} and potentially large magnitude of the orbital Hall effect~\cite{kontani2008,kontani2009,go2018,jo2018} have attracted the community to the field of \textit{orbitronics}~\cite{bernevig2005b,phong2019,go2017,Go2021}. 
	Nevertheless, experimental detection of orbital dynamics~\cite{stamm2019,ding2020,zheng2020,tazaki2020,kim2021,lee2021,ding2021} remains indirect because orbital and spin signals are hard to distinguish. This distinction remains challenging even theoretically ~\cite{go2020,ko2020,xiao2020,cysne2021,sahu2021,bhowal2021} because most dynamics of orbital ($\vec{L}$) and spin ($\vec{S}$) are rooted in the same commutator algebra, $[L_\alpha,L_\beta]\propto i\varepsilon_{\alpha\beta\gamma}L_\gamma$ and $[S_\alpha,S_\beta]\propto i\varepsilon_{\alpha\beta\gamma}S_\gamma$. 
	
	In this Letter, we reveal a fundamental difference between the orbital and spin degrees of freedom. This difference arises because the three OAM operators $L_\alpha$ are incomplete and additional operators are necessary to describe the orbital dynamics completely. The additional operators include even-order symmetrized products of $L_\alpha$, which we call the \textit{orbital angular position} (OAP) operators, such as $\{L_\alpha,L_\beta\}$. In contrast, the three spin operators $S_\alpha$ provide a complete description of the spin dynamics. This difference gives rise to orbital dynamics, which do not have spin counterparts and thus may be used for the distinction. We construct a general Hamiltonian of orbital systems with the time-reversal (TR) and inversion symmetries, and examine various linear responses of the orbital dynamics to external electric fields or chemical potential gradients. Using the quantum Boltzmann approach, we derive orbital drift-diffusion equations including various orbital transport coefficients mediated by the OAM and the OAP.
	
	The intrinsic orbital and spin Hall effects in transition metals~\cite{tanaka2008,kontani2008} are good examples to illustrate the difference between the orbital and spin degrees of freedom. The Hall effects require ${\bf L}$ and ${\bf S}$ to be coupled to the crystal momentum ${\bf k}$. When both TR and inversion symmetries are present as in transition metals, the ${\bf S}$-${\bf k}$ coupling is forbidden but the ${\bf L}$-${\bf k}$ coupling {\it is} possible. For instance, the coupling $k_\alpha k_\beta (L_\gamma L_\chi + L_\chi L_\gamma)$ is compatible with both symmetries, but its spin counterpart $k_\alpha k_\beta (S_\gamma S_\chi + S_\chi S_\gamma)$ is not possible since the spin operator algebra $S_\gamma S_\chi + S_\chi S_\gamma=\{ S_\gamma,S_\chi \}\propto \delta_{\gamma\chi}$ makes the ${\bf S}$-${\bf k}$ coupling \textit{independent} of ${\bf S}$. Because of this difference, the orbital Hall effect can arise purely from the orbital degree of freedom itself whereas the spin Hall effect should be mediated by a different degree of freedom. In transition metals~\cite{tanaka2008,kontani2008}, ${\bf L}$ mediates the effective ${\bf S}$-${\bf k}$ coupling through the ${\bf L}$-${\bf k}$ coupling and the spin-orbit coupling ${\bf S}\cdot{\bf L}$. Note that the OAP operator $\{L_\gamma,L_\chi\}$ characterizes the crucial difference between the orbital and spin degrees of freedom.

	{\it Effective Hamiltonian of orbital systems.---} To focus on the orbital dynamics, we assume the spin-orbit coupling to be absent and ignore the spin degree of freedom. Then, the effective Hamiltonian of orbital systems with both TR and inversion symmetries can be constructed from symmetry-compatible ${\bf L}$-${\bf k}$ couplings, $k_{\alpha_1}k_{\alpha_2}\cdots k_{\alpha_{m}}(L_{\beta_1}L_{\beta_2}\cdots L_{\beta_n}+{\rm h.c.})$, where $m$ and $n$ are even integers. Considering up to $n=2$, one obtains the effective Hamiltonian,
	\begin{equation}
		H(\vec{k})= h_0(\vec{k})+\sum_{\alpha,\beta=x,y,z} h_{2,\alpha\beta}(\vec{k}) \{ L_\alpha,L_\beta \}, \label{effectivehamiltonian}
	\end{equation}
	where $h_0({\bf k})$ and $h_{2,ab}({\bf k})$ are even in $\vec{k}$. 
	Note that $H(\vec{k})$ is a generalized version of the Hamiltonian considered in Ref.~\cite{bernevig2005b}. Here we focus on $p$-orbital systems, for which Eq.~(\ref{effectivehamiltonian}) turns out to be the most general Hamiltonian since couplings with larger $n$ can be expressed as linear combinations of the couplings with $n\le 2$~\cite{supple}. Generalization of $H({\bf k})$ to arbitrary orbitals (e.g. $d,f$ orbitals) is also possible~\cite{supple}.  The $\vec{k}$ dependence of $h_{2,\alpha\beta}(\vec{k})$ describes the orbital texture. Since it arises from the orbital-dependent electron hopping and the crystal field, the energy scale of $h_{2,\alpha\beta}({\bf k})/\hbar^2$ is of the order of a few eVs. Thus the ${\bf L}$-${\bf k}$ coupling is much stronger than the spin-orbit coupling.
	
	Now we illustrate the physical meaning of the OAP operator $\{L_\alpha,L_\beta\}$ in Eq.~(\ref{effectivehamiltonian}). 
	When $\beta=\alpha$, simple algebra shows $\{ L_{\alpha},L_{\alpha} \}= 2\hbar^2(1-|p_{\alpha}\rangle \langle p_{\alpha}|)$. Thus, the diagonal OAP operator $\{ L_{\alpha},L_{\alpha} \}$ measures the \textit{orbital polarization} for the $p_{\alpha}$ state. For $\beta\neq \alpha$, consider an orbital state $|u\rangle$ obtained by rotating $|p_\alpha\rangle$ towards the $\beta$ direction,
	{\it i.e.} $|u\rangle=\cos\phi |p_{\alpha}\rangle+\sin\phi |p_{\beta}\rangle$ ($\langle p_\alpha|p_\beta \rangle=0$). Simple algebra shows $\langle u|\{L_{\alpha},L_{\beta}\}|u\rangle=-\sin 2\phi$, implying that the off-diagonal OAP operator $\{L_{\alpha},L_{\beta}\}$ measures the \textit{orbital torsion} away from the $\alpha$ axis (for $\sin\phi\approx 0$) or the $\beta$ axis (for $\cos\phi\approx 0$) around the axis perpendicular to the $\alpha\beta$ plane.
	
\begin{figure}
		\includegraphics[width=8.6cm]{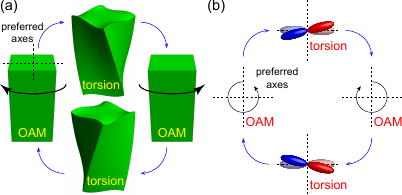}
		\caption{\label{Fig:orbital torsion} Analogy between (a) the classical torsion pendulum and (b) the orbital Hanle oscillation. The elastic beams in this figure were generated by modifying the source code at Ref.~\cite{torsion}.}
	\end{figure}

	{\it Orbital dynamics.--}
	When $\vec{k}$ is conserved, one obtains from Eq.~\eqref{effectivehamiltonian} the following equations of motion for each ${\bf k}$~\cite{supple}:
	\begin{subequations}
		\label{precession}
		\begin{gather}
			\frac{dL_\alpha}{dt}=
			\frac{[L_\alpha,H(\vec{k})]}{i\hbar}=
			\sum_{\beta\gamma}A_{\alpha\beta\gamma}({\bf k}) \{L_\beta,L_\gamma\},
			\label{precession_a}\\
			\frac{d\{L_\alpha,L_\beta\}}{dt}
			=\frac{[\{L_\alpha,L_\beta\},H(\vec{k})]}{i\hbar}
			=\sum_\gamma B_{\alpha\beta\gamma}(\vec{k})L_\gamma,
			\label{precession_b}
		\end{gather}
	\end{subequations}
	where real coefficients $A_{\alpha\beta\gamma}(\vec{k})$ and $B_{\alpha\beta\gamma}(\vec{k})$ are combinations of $h_{2,\alpha\beta}(\vec{k})$. 
	The combination of the OAP operators on the right-hand side of Eq.~\eqref{precession_a} measures the orbital torsion away from the eigenorbital direction of $H({\bf k})$ around the $\alpha$ direction. This orbital torsion induces the OAM $L_\alpha$ 
	to change [Eq.~\eqref{precession_a}] and the OAM induces the orbital torsion to change [Eq.~\eqref{precession_b}]. 
	Thus, Eq.~\eqref{precession} resembles the equations of motion for a classical torsion pendulum (Fig.~\ref{Fig:orbital torsion}).
	To illustrate the orbital dynamics, suppose that eigenstate wave functions at a particular momentum $\vec{k}_0$ have $p_{x'}$, $p_{y'}$, $p_{z'}$ characters with energies $E_{p_{x'}}$, $E_{p_{y'}}$, $E_{p_{z'}}$, respectively, where $x',y',z'$ are mutually orthogonal directions. $H(\vec{k}_0)$ is then given by $\sum_{\alpha'} E_{p_{\alpha'}} |p_{\alpha'}\rangle \langle p_{\alpha'}|$.
	Then for a nonequilibrium state at $t=0$, $\ket{u(t=0)}_{\rm neq}=\cos\phi\ket{p_{x'}}+\sin\phi\ket{p_{y'}}$ at the momentum ${\bf k}_0$,
	its initial expectation values $\langle L_{z'}(t=0) \rangle=0$ and $\langle \{ L_{x'},L_{y'} \}(t=0) \rangle=-\sin 2\phi$ evolve to $\langle L_{z'}(t) \rangle=-\sin 2\phi \sin \omega t$ and $\langle \{ L_{x'},L_{y'} \}(t) \rangle=-\sin 2\phi \cos \omega t$, where $\omega=(E_{p_{x'}}-E_{p_{y'}})/\hbar$. Note that the oscillation between the OAM $L_{z'}$ and the orbital torsion $\{L_{x'},L_{y'}\}$ does not require any symmetry breaking, which is in clear contrast to the spin oscillation that requires either the TR or the inversion symmetries to be broken.
	
	{\it Quantum Boltzmann formalism.--} 
	When $\vec{k}$ is not conserved due to an external electric field $\vec{\mathcal{E}}$ 
	or chemical potential gradients, the OAM and the OAP dynamics couple to the $\vec{k}$ dynamics. %
	Then, due to the orbital texture~\cite{go2018} [or the ${\bf k}$ dependence of $h_{2,\alpha\beta({\bf k})}$ in Eq.~\eqref{effectivehamiltonian}], the ${\bf k}$ dynamics naturally generates nonequilibrium superpositions and results in the orbital Hall effect~\cite{kontani2008,kontani2009,go2018,jo2018,bernevig2005b,tokatly2010} despite the strong tendency of the crystal field to suppress the OAM. We use the quantum Boltzmann formalism to investigate the coupled dynamics of the OAM, the OAP, and ${\bf k}$ systematically, and unveil other interesting orbital dynamics that were not recognized in previous studies~\cite{kontani2008,kontani2009,go2018,jo2018,bernevig2005b,tokatly2010}.
	
	To obtain a complete picture of the orbital dynamics, We start from the linearized quantum Boltzmann equation~\cite{kim2020}  $-(e\vec{\cal E}/\hbar)\cdot \nabla_{\bf k} f^0 +(1/2) \{\textbf{v};\nabla_{\bf r}g\}-(1/i\hbar)[H,g] = \langle \partial_tf\rangle_{\rm imp}$ for the steady-state Wigner distribution function $f(\vec{r},\vec{k})=f^0(\vec{r},\vec{k})+g(\vec{r},\vec{k})$,
	where $\vec{r}$ is the position, $f^0$ is the equilibrium Fermi-Dirac distribution, $g$ is the nonequilibrium deviation, $-e<0$ is the electron charge, $\vec{v}=(1/\hbar)\partial_\vec{k}H$ is the velocity operator, $\{ \cdots ; \cdots \}$ is the inner-product anticommutator, $\langle \partial_tf\rangle_{\rm imp}$ is the impurity contribution. 
	For complete description of orbital states in $p$-orbital systems, one should deal with a $3\times 3$ matrix $f$, which contains 9 independent information (charge, three components of $\vec{L}$, and five components of $\{L_\alpha,L_\beta\}$~\cite{note1}). Thus, the OAP is an indispensable element of the Boltzmann analysis.
	In contrast, the Boltzmann equation analysis for spin dynamics~\cite{kim2020,haney2013} usually deals with a $2\times 2$ matrix $f$, which contains 4 independent information (charge and three components of $\vec{S}$). Thus it is evident that the OAP is the source of the difference between the orbital and spin dynamics. 
	
	We decompose $g$ into isotropic and anisotropic parts, $g=\sum_{mn}\left[ g^{\rm iso}_{mn}(\vec{k})\ket{m\vec{k}}\bra{n\vec{k}} + g^{\rm an}_{mn}(\vec{k})\ket{m\vec{k}}\bra{n\vec{k}}\right]$, where $g^{\rm iso}(\vec{k})$ is the direction-independent part, $g^{\rm an}(\vec{k})$ is the angle harmonic part, 
	and $|m{\bf k}\rangle$, $|n{\bf k}\rangle$ are eigenstates in the bands $m,n$. 
	We consider up to first-harmonic anisotropy~\cite{valet1993}. Then the quantum Boltzmann equation is decomposed into 
	\begin{subequations}
		\label{QBEs}
		\begin{gather}
			\frac{1}{2} \{\textbf{v};\nabla_{\bf r}g^{\rm an}\}-\frac{[H,g^{\rm iso}]}{i\hbar} =\langle\partial_tf\rangle^{\rm iso}_{\rm imp}, \label{isoeq} \\ 
			-\frac{e\vec{\cal E}}{\hbar}\cdot \nabla_\vec{k} f^0 +\frac{1}{2} \{\textbf{v};\nabla_{\bf r}g^{\rm iso}\}-\frac{[H,g^{\rm an}]}{i\hbar} =\langle\partial_tf\rangle^{\rm an}_{\rm imp}. \label{anieqn}
		\end{gather}
	\end{subequations}
	The isotropic part $g^{\rm iso}$ is related to the chemical potential $\mu$ by
	$g^{\rm iso}=f^0(\mu+E_F-H)-f^0(E_F-H)$, where $E_F$ is the Fermi level~\cite{kim2020} and the $3\times 3$ matrix $\mu$ can be expanded by $\mu=\sum_\alpha \mu_\alpha L_\alpha+\sum_{\alpha\beta} \mu_{\alpha\beta} \{L_\alpha,L_\beta\}$. Here, $\mu_\alpha$ is the OAM chemical potential and $\mu_{\alpha\beta}$ is the OAP chemical potential. The anisotropic part $g^{\rm an}$ gives rise to the OAM current $-(e/2V)\Tr[\{L_\alpha,\vec{v}\}g^{\rm an}]$ and the OAP current $-(e/2V)\Tr[\{\{L_\alpha,L_\beta\},\vec{v}\}g^{\rm an}]$. Then, their transport equations are given by Eq.~(\ref{QBEs}) expanded by the 9 independent bases. 
	
	{\it Example: Two-dimensional $p$-orbital system.--} Three-dimensional $p$-orbital systems allow 9 types of currents and 9 types of chemical potentials. Such a large number of options make the systems less illustrative. To reduce the number of options, we take a simpler two-dimensional $p$-orbital system in the $xy$ plane with the $p_z$ orbital discarded under the assumption that states near the Fermi energy have only $p_x$ and $p_y$ characters. Then, there are only 4 types of currents and 4 types of chemical potentials, just like 
	the spin dynamics. Nevertheless, this orbital system still retains crucial differences from spin systems as demonstrated below and is thus an illustrative system with minimal technical complexity.
	
\begin{figure}
		\includegraphics[width=8.6cm]{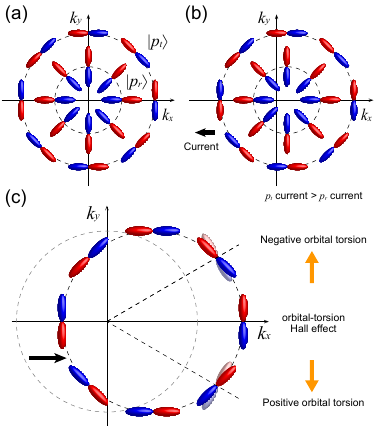}
		\caption{\label{Fig:OTHE}Illustrations of (a) the orbital texture in Eq.~(\ref{2D_hamiltonian}), (b) orbital-polarization conductivity ($\chi_{xx}^3$), and (c) the orbital-torsion Hall conductivity ($\chi_{yx}^1$). Note that electrons with positive or negative $v_y$ (denoted by yellow arrows) have negative or positive orbital torsion.}
	\end{figure}

	As a concrete system, we take the following Hamiltonian:
	\begin{equation}
		H(\vec{k})=\frac{\hbar^2 k^2}{2m_r^*}\ket{p_r(\vec{k})}\bra{p_r(\vec{k})}
		+ \frac{\hbar^2 k^2}{2m_t^*}\ket{p_t(\vec{k})}\bra{p_t(\vec{k})}, 
		\label{2D_hamiltonian}
	\end{equation}
	whose energy eigenstates are characterized by radial orbital states $\ket{p_r(\vec{k})}=\cos\phi_\vec{k}\ket{p_x}+\sin\phi_\vec{k}\ket{p_y}$ and tangential orbital states $\ket{p_t(\vec{k})}=-\sin\phi_\vec{k}\ket{p_x}+\cos\phi_\vec{k}\ket{p_y}$ with $\phi_\vec{k}=\arg(k_x+ik_y)$. $H({\bf k})$ has the orbital texture [Fig.~\ref{Fig:OTHE}(a)] required for the orbital Hall effect~\cite{go2018}. We note that $H(\vec{k})$ is identical to the effective Hamiltonian of a $p$-doped graphane~\cite{tokatly2010,supple}, whose topmost valence bands have mainly $p_x$ and $p_y$ characters~\cite{cudazzo2010}. Actually, $H(\vec{k})$ is applicable to general two-orbital systems with approximate rotational symmetry near the $\Gamma$ point. Because of the difference between $\sigma$ and $\pi$ hopping integrals or due to the $sp$ hybridization, the two effective masses $m_r^*$, $m_t^*$ are generally different~\cite{supple}. Their difference is parametrized by the dimensionless parameter $\eta$ and the harmonic mean mass $m^*$ as $1/m_{r/t}^*=(1\pm\eta)/m^*$.
	
	To derive the drift-diffusion equations, we define various chemical potentials and currents. 
	The $2\times 2$ matrix chemical potential is expanded as $\mu=\sum_i \mu^{(i)}\sigma_i'$, where $\sigma_i'$ are the Pauli matrices in the eigenstate basis: $\sigma_0'=I$, $\sigma_1'=\ket{p_r}\bra{p_t}+\ket{p_t}\bra{p_r}$, $\sigma_2'=-i\ket{p_r}\bra{p_t}+i\ket{p_t}\bra{p_r}$, and $\sigma_3'=\ket{p_r}\bra{p_r}-\ket{p_t}\bra{p_t}$~\cite{note2,note3}. With these definitions, $\mu^{(i)}$'s for $i=0,1,2,3$ are interpreted as the charge, orbital torsion, OAM, and orbital-polarization chemical potentials, respectively. Similarly, the current is defined as
	\begin{equation}
		\vec{J}^{(i)}=-\frac{e}{2\hbar V}\sum_\vec{k}\Tr[\{\partial_\vec{k}H,\sigma_i'\}g^{\rm an}],\label{current}
	\end{equation}
	where $i=0,1,2,3$. For further simplification, we adopt the relaxation time approximation~\cite{kim2020}:
	$
	\langle \partial_tf\rangle_{\rm imp} =
	-(g^{\rm iso}_1 \sigma_1' + g^{\rm iso}_2\sigma_2')/\tau_{\rm dp}-g^{\rm iso}_3\sigma_3'/\tau_{\rm of}-g^{\rm an}/\tau_m,
	$
	where $g_i^{\rm iso}=(1/2)\Tr[\sigma_i'g^{\rm iso}]$. $\tau_{\rm dp}$ is the dephasing time of coherent superpositions, $\tau_{\rm of}$ is the orbital-flip time between the eigenstates, and $\tau_m$ is the momentum scattering time. 
	
	From Eqs.~(\ref{anieqn}), one can express $g^{\rm an}$ in terms of $\nabla_\vec{r}g^{\rm iso}(\propto \nabla_\vec{r}\mu^{(i)})$ and the electric field $\boldsymbol{\mathcal{E}}$. Plugging $g^{\rm an}$ into Eq.~(\ref{current}) gives the linear response coefficients as
	\begin{equation}
		J_\alpha^{(i)} = \sum_{\beta} \chi^i_{\alpha\beta}\mathcal{E}_\beta + \sum_{j\alpha\beta} \xi^{ij}_{\alpha\beta} \nabla_\beta \mu^{(j)}. \label{currentresult}
	\end{equation}
	Geometrical symmetries force many components of the conductivity tensor $\chi_{\alpha\beta}^{i}$ and the diffusivity tensor $\xi_{\alpha\beta}^{ij}$ to vanish. Because of the rotation symmetry of $H(\vec{k})$, the drift-diffusion dynamics is isotropic and we may set $\beta=x$ [Eq.~\eqref{currentresult}] without loss of generality. Then, the conductivity tensor $\chi_{\alpha x}^{i}$ can have only four nonvanishing components. Two of them are longitudinal components, the Drude charge conductivity $\chi_{xx}^{0}$ and the orbital-polarization conductivity $\chi_{xx}^{3}$. The latter quantifies the difference between the $p_r$- and $p_t$-polarized longitudinal 
	currents arising from the $x$-shifted inner ($p_r$-polarized) and outer ($p_t$-polarized) Fermi surfaces, respectively [Fig.~\ref{Fig:OTHE}(b)]. Thus, the longitudinal electron flow in {\it nonmagnetic} systems is partially {\it orbital polarized} just as the longitudinal electron flow in ferromagnets is partially spin polarized.
	The other two are Hall components, the orbital Hall conductivity $\chi_{yx}^2$~\cite{bernevig2005b,tanaka2008,kontani2008,kontani2009,go2018} and the orbital-torsion Hall conductivity $\chi_{yx}^1$. The latter quantifies the difference between the $(p_r+p_t)$- and $(p_r-p_t)$-polarized Hall currents arising from the $x$-shifted Fermi surfaces [Fig.~\ref{Fig:OTHE}(c)]. The orbital-torsion Hall effect has not been  recognized before but its presence is natural considering the mutual induction of the OAM and the orbital torsion [Eq.~(\ref{precession})]. 
	
	\begin{figure}
		\includegraphics[width=8.6cm]{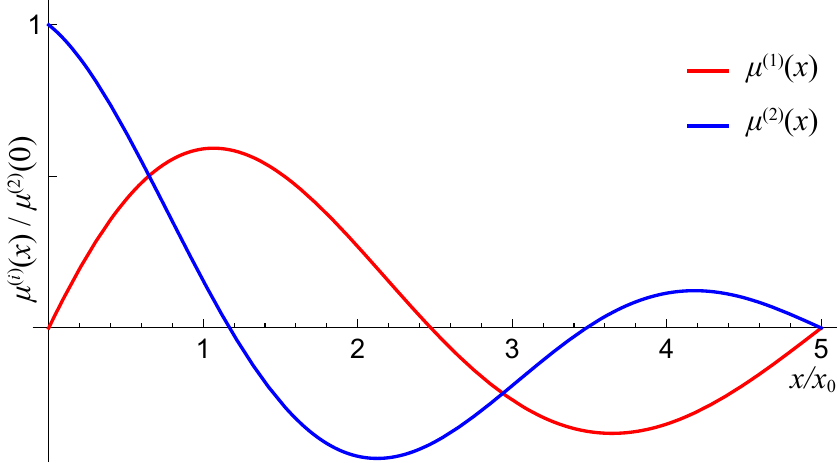}
		\caption{\label{Fig:solution} Solution of Eq.~\eqref{orbital DDE12} with the boundary conditions $\mu^{(1)}(0)=\mu^{(1)}(L)=\mu^{(2)}(L)=0$ and $\mu^{(2)}(0)\neq 0$, which simulate the case where an OAM current is injected from the left boundary ($x=0$) and reflected at the other boundary ($x=L$).	The following parameters are used: $\eta=0.3$, $\tau_{\rm dp}/\tau_m=5$, $E_F\tau_m/\hbar=5$, and $L=5x_0$. Here, $x_0=\sqrt{\hbar\tau_m/m^*} \approx 1~\rm nm$ for $\tau_m=10~\rm fs$ and $m^*=9.1\times10^{-31}~\mathrm{kg}$.
		}
	\end{figure}
	
	The diffusivity tensor $\xi_{\alpha x}^{ij}$ can have 16 nonvanishing components. Half of them are the diffusivities for the charge and orbital-polarization chemical potential gradients ($\xi_{\alpha x}^{i0}$ and $\xi_{\alpha x}^{i3}$). They possess similar phenomenology to $\chi_{\alpha x}^{i}$ as inferred from the Einstein relation~\cite{datta1995,song2020}.
	The other half are the diffusivities for the OAM and the orbital-torsion chemical potential gradients ($\xi_{\alpha x}^{i2}$ and $\xi_{\alpha x}^{i1}$). For longitudinal diffusion, there are four such components: $\xi_{xx}^{22}$ (OAM diffusivity), $\xi_{xx}^{11}$ (orbital-torsion diffusivity), $\xi_{xx}^{21}$, and $\xi_{xx}^{12}$. The former two can be easily understood but the latter two are unanticipated. $\xi_{xx}^{21}$ implies that the orbital-torsion chemical potential gradient, which may be interpreted as {\it real}-space orbital texture, generates the {\it longitudinal} OAM current. %
	$\xi_{xx}^{12}$ implies the opposite of $\xi_{xx}^{21}$. Both $\xi_{xx}^{21}$ and $\xi_{xx}^{12}$ are due to the mutual induction of the OAM and the orbital torsion [Eq.~\eqref{precession}] and do not have spin counterparts. For Hall diffusion, there are again four components: $\xi_{yx}^{02}$, $\xi_{yx}^{32}$, $\xi_{yx}^{01}$, and $\xi_{yx}^{31}$. The former two amount to the inverse orbital Hall effect and the associated orbital-polarization current generation. The latter two amount to the inverse orbital-torsion Hall effect and the associated orbital-polarization current generation. 
	
	%
	{\it Orbital diffusion equations.--}
	To obtain the orbital diffusion equations, we trace out Eq.~(\ref{isoeq}) over $\vec{k}$ to obtain a $2\times 2$ matrix continuity equation. Its four components can be reduced, by using Eq.~(\ref{currentresult}), 
	to the closed orbital diffusion equations: 
	\begin{subequations}
		\label{orbital DDE}
		\begin{gather}
			\nabla^2 \mu^{(0)}= 0,~\nabla^2\mu^{(3)}= \frac{\mu^{(3)}}{\lambda_{\rm of}^2},\label{orbital DDE03}\\
			\Lambda^2\nabla^2 \begin{pmatrix}
				\mu^{(1)}\\
				\mu^{(2)}
			\end{pmatrix} = \begin{pmatrix}
				\mu^{(1)}\\
				\mu^{(2)}
			\end{pmatrix},\label{orbital DDE12}
		\end{gather}
	\end{subequations}
	where 
	$\lambda_{\rm of}=\hbar k_F\sqrt{\tau_{\rm of}\tau_m}/\sqrt{2}m^*$, 
	$k_F=\sqrt{2m^*E_F}/\hbar$, respectively. $\Lambda^2$ is a $2\times2$ matrix whose explicit expression is presented in Ref.~\cite{supple}. 
	The first equation in Eq.~(\ref{orbital DDE03}) implies the charge conservation and is trivial. The second equation in Eq.~(\ref{orbital DDE03}) has the same form as the spin chemical potential relaxation equation~\cite{valet1993}, implying that the orbital counterpart of the spin chemical potential is the orbital-polarization chemical potential $\mu^{(3)}$ rather than
	the OAM chemical potential $\mu^{(2)}$. The latter, on the other hand, is coupled to the orbital-torsion chemical potential $\mu^{(1)}$ through Eq.~\eqref{orbital DDE12}. When the chemical potentials vary only along one direction (say $x$), the general solution of Eq.~(\ref{orbital DDE12}) is given by a linear combination of $e^{\pm x/\lambda_1}$ and $e^{\pm x/\lambda_2}$ where $\lambda_i^2$ are the eigenvalues of $\Lambda^2$, which turn out to be complex~\cite{supple} for the model system [Eq.~\eqref{2D_hamiltonian}].  Therefore, $\mu^{(1)}$ and $\mu^{(2)}$ exhibit oscillatory decay (Fig.~\ref{Fig:solution}). 
	Its oscillation wavelength is proportional to $1/(k_{F,r}-k_{F,t})$, where $k_{F,r/t}$ is the Fermi wave vector for the radial/tangential orbital bands. Hence the nature of the oscillatory decay is the dephased oscillation, similar to the oscillatory decay of noncollinear spins in ferromagnets~\cite{stiles2002}. Because of the dephasing nature, the oscillation may be suppressed and monotonic decay may appear instead when energy spacing between energy bands varies with ${\bf k}$ significantly unlike the model system [Eq.~\eqref{2D_hamiltonian}] with isotropic spacing. We emphasize, however, that the OAM oscillation does not require any symmetry breaking in contrast to the spin oscillation that requires  either the TR~\cite{johnson1985,jedema2002} or the inversion symmetry~\cite{koo2009,manchon2015} to be broken. 
	
	\begin{figure}
		\includegraphics[width=8.6cm]{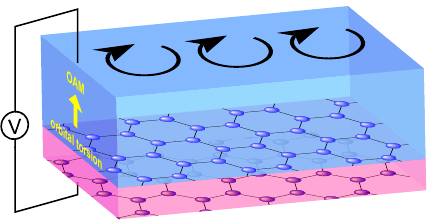}
		\caption{\label{Fig:detection} Twistronic detection of orbital torsion transport.
		}
	\end{figure}
	
	{\it Discussion and outlook.--} We propose a few experiments to probe unique characteristics of the orbital dynamics. A laser may generate states with nonzero OAM or nonzero orbital torsion at a selected $\vec{k}$, which oscillates in time [Eq.~\eqref{precession}] even after the light is turned off. Measuring such oscillation with, e.g., time-resolved magneto-optic Kerr effect (TR-MOKE) would enable experimental detection of the orbital dynamics~\cite{note4}.
	$\xi_{xx}^{21}$ motivates experiments with twisted heterostructures as shown in Fig.~\ref{Fig:detection}. When a bias is applied perpendicularly to a twisted heterostructure, the twist may play the role of the orbital-torsion chemical potential gradient along the out-of-plane direction and generate an out-of-plane OAM current, which can be detected by means of the orbital torque~\cite{go2020} or magnetoresistance~\cite{ko2020}. Since the bias-induced orbital torsion may affect the hopping integrals, measuring transport properties as a function of the gate bias would enable examining the effects of the nonequilibrium orbital torsion. For candidate materials for Fig.~\ref{Fig:detection}, we propose nonmagnetic van der Waals materials with small spin-orbit coupling, such as $\rm CrS_2$ and $\rm MoS_2$. Furthermore, when combined with the spin-orbit coupling, the OAM current is accompanied by a spin current, which may shed light on the mysterious long-distance spin transport in chiral systems~\cite{gohler2011,xie2011,suda2019,shiota2021,waldeck2021} since the heterostructure in Fig.~\ref{Fig:detection} amounts to an artificial version of chiral crystals.
	Combining orbitronics and twistronics~\cite{andrei2020} may expand the scope of the orbital physics significantly.
	
	A few remarks are in order. First, more diverse orbital dynamics may emerge when more than two orbitals are involved or in three dimensions. Still more interesting dynamics may emerge when the spin-orbit coupling is strong and a system becomes effectively a higher orbital systems with the total angular momentum $\vec{J}=\vec{L}+\vec{S}$. Second, during the conversion between the OAM and the orbital torsion, the angular momentum is instantaneously transferred to lattice. Therefore, simultaneous dynamics of the OAM, the spin, and the phonon angular momentum~\cite{hamada2018,zhang2014,park2020} may be required for more complete study of the angular momentum flow. 
	
	\begin{acknowledgments}
		We acknowledge S. B. Chung, S.-H. Rhim, and K.-J. Lee for discussions. H.-W. L acknowledges B.-C. Min for fruitful discussion and hospitality during his sabbatical visit at the Korea Institute of Science and Technology (KIST). S.H. and H.-W. L were supported by the Samsung Science and Technology Foundation (BA-1501-51). K.-W. K acknowledges the financial support from the KIST Institutional Programs (2E31541, 2E31542), the National Research Foundation (NRF) of Korea (2020R1C1C1012664), and the National Research Council of Science and Technology (NST) (CAP-16-01-KIST).
	\end{acknowledgments}



\section*{Supplementary material}

# Supplemental Material for "Orbital Dynamics in Centrosymmetric Systems"

Seungyun Han,[1] Hyun-Woo Lee,[1, *] and Kyoung-Whan Kim[2, †]

[1]*Department of Physics, Pohang University of Science and Technology, Pohang 37673,Korea*
[2]*Center for Spintronics, Korea Institute of Science and Technology , Seoul 02792, Korea*

## I. GENERAL ORBITAL HAMILTONIAN AND THE HEISENBERG EQUATION

### A. $p$ orbitals

Here we prove that Eq. (1) in the main text is the most general Hamiltonian for $p$ orbitals. For $p$ orbitals, the orbital angular momentm (OAM) operators are $3 \times 3$ matrices. From their explicit expressions, it is easy to show that *any* $3 \times 3$ matrix $A$ can be expanded by

$$A = \sum_\alpha a_\alpha L_\alpha + \sum_{\alpha\beta} b_{\alpha\beta}\{L_\alpha, L_\beta\}, \tag{S1a}$$

where

$$a_\alpha = \frac{1}{2\hbar} \operatorname{Tr}[AL_\alpha], \ b_{\alpha\beta} = \frac{1+\delta_{\alpha\beta}}{2\hbar^2} \operatorname{Tr}[A\{L_\alpha, L_\beta\}] - \frac{\delta_{\alpha\beta}}{2} \operatorname{Tr}[A]. \tag{S1b}$$

Therefore, any higher-order product can be represented as linear combinations of first- and second-order products. Note that the number of first-order products and that of second-order products are respectively 3 and 6, whose sum matches with $9 = 3 \times 3$. This implies that they form a complete basis for $3 \times 3$ matrices since they are linearly independent. The Hamiltonian [Eq. (1)] is obtained by discarding the first-order products by symmetries and using $I = (L_x^2 + L_y^2 + L_z^2)/2\hbar^2$ to extract $h_0(\mathbf{k})$.

Although the equation of motion [Eq. (2)] can be obtained by direct calculations, for later purpose, we derive it by using the time-reversal (TR) property of $\mathbf{L}$ only. Note that $H(\mathbf{k})$ is given by second-order products. To express $dL_\alpha/dt$, we need to calculate $(1/i\hbar)[L_\alpha, \{L_\beta, L_\gamma\}]$. From Eq. (S1a), we can expand it to $(1/i\hbar)[L_\alpha, \{L_\beta, L_\gamma\}] = \sum_\delta a_\delta L_\gamma + \sum_{\delta\epsilon} b_{\delta\epsilon}\{L_\delta, L_\epsilon\}$, where $a_\delta = (1/2i\hbar^2) \operatorname{Tr}[L_\delta[L_\alpha, \{L_\beta, L_\gamma\}]]$. Note that, by taking TR operation ($i \to -i$ and $\mathbf{L} \to -\mathbf{L}$) $a_\delta$ transforms to $-a_\delta$, implying that $a_\delta$ is purely imaginary. Now we take the complex conjugate of $a_\delta$.

$$a_\delta^* = -\frac{1}{2i\hbar^2} \operatorname{Tr}[L_\delta[L_\alpha, \{L_\beta, L_\gamma\}]]^\dagger = \frac{1}{2i\hbar^2} \operatorname{Tr}[L_\delta[L_\alpha, \{L_\beta, L_\gamma\}]] = a_\delta. \tag{S2}$$

At the second step, we use the fact that a commutator (anticommutator) of Hermitian matrices is anti-Hermitian (Hermitian). Equation (S2) implies that $a_\delta$ is purely real. Therefore, $a_\delta$ should be zero. Since $(1/i\hbar)[L_\alpha, \{L_\beta, L_\gamma\}]$ has no first-order product element, one obtains Eq. (2a). The same logic applies to the derivation of Eq. (2b).

### B. Generalization for arbitrary orbital quantum numbers

Now we generalize the above derivations for $p$ orbital to arbitrary orbital systems with angular momentum quantum number $l \geq 1$. First we define the symmetrized product, which is given by an average over all possible permutations of products, e.g., $\operatorname{Symm}(L_x L_y) = \{L_x, L_y\}/2$ and $\operatorname{Symm}(L_x L_y^2) = (L_x L_y^2 + L_y L_x L_y + L_y^2 L_x)/3$. To express a general Hamiltonian, it is necessary to find a complete set of bases spanning all possible $(2l+1) \times (2l+1)$ matrices. One can show that the $(2l-1)$-order and $2l$-order symmetrized products of $L_\alpha$ provide a complete basis set of $(2l+1) \times (2l+1)$ matrices. Note that the number of possible combinations of $(2l-1)$-order symmetrized products is $l(2l+1)$ and that of $2l$-order symmetrized products is $(l+1)(2l+1)$, so their sum matches with $(2l+1) \times (2l+1)$. Therefore, a general Hamiltonian is given by

$$H(\mathbf{k}) = \sum_{\alpha_1 \cdots \alpha_n} h_{n,\alpha_1 \cdots \alpha_n}(\mathbf{k}) \operatorname{Symm}(L_{\alpha_1} \cdots L_{\alpha_n}), \tag{S3}$$

where $n = 2\lfloor l \rfloor$, where $\lfloor \cdots \rfloor$ is the floor function. For example, $l = 2$ ($d$ orbitals), the general Hamiltonian is given by a summation of fourth-order symmetrized products: $H(\mathbf{k}) = \sum_{\alpha\beta\gamma\delta} h_{4,\alpha\beta\gamma\delta}(\mathbf{k}) \operatorname{Symm}(L_\alpha L_\beta L_\gamma L_\delta)$. We emphasize that the

---

* hwl@postech.ac.kr
† kwk@kist.re.kr

Hamiltonian is still written as a linear combination of the (generalized) orbital torsion operators. When TR or centrosymmetry is broken, one may add odd-order products. Note that for spin ($l = 1/2$), the Hamiltonian includes only the zeroth order products (i.e., identity), making the case for $l = 1/2$ and that for $l > 1/2$ significantly different.

The time evolution of the $(2l − 1)$-order and the $2l$-order products are related in the same way as Eq. (2) and show oscillatory behaviors. One can directly apply the same logic in Eq. (S2) by using TR properties of $\mathbf{L}$. In other words,

$$\frac{d}{dt}[2l\text{-order products})] = \sum[(2l − 1)\text{-order products}], \tag{S4a}$$

$$\frac{d}{dt}[(2l − 1)\text{-order products})] = \sum[2l\text{-order products}]. \tag{S4b}$$

Therefore, the oscillation between TR-odd excitation and TR-even excitation demonstrated in the main text remains the same. Here the odd-order product generalizes the OAM and the even-order product generalizes the orbital torsion. However, an odd-order product may correspond to not only nonzero OAM states but also some zero OAM states with TR-odd properties. For instance, in $d$ orbital, imaginary mixture of $e_g$ orbitals, e.g., $|d_{3z^2-r^2}\rangle + i |d_{x^2-y^2}\rangle$ has no OAM but odd in TR. Indeed, this state can be measured by the third-order product $\text{Symm}(L_x L_y L_z)$. We note that the imaginary mixture of $e_g$ orbitals was experimentally measured by Ref. [56]

## II. MICROSCOPIC DERIVATION OF EQ. (4)

In this section, we consider a two-dimensional hexagonal lattice system with $s$ and $p$ orbitals as an example system to which $H(\mathbf{k})$ in Eq. (4) becomes relevant. To be specific, we demonstrate that the effective Hamiltonian for $p_x$- and $p_y$-character bands in this system can be described by $H(\mathbf{k})$. We suppose this system has the mirror symmetry, $z \to -z$. Then $p_z$ orbitals are decoupled from $s$, $p_x$, and $p_y$ orbitals. Thus $p_z$ orbitals may be ignored for the purpose of deriving an effective Hamiltonian for $p_x$ and $p_y$ orbitals. To examine the coupled dynamics of two $s$ orbitals, two $p_x$ orbitals, and two $p_y$ orbitals (recall that the hexagonal lattice contains two atoms in a unit cell), we start with the following $6 \times 6$ tight-binding Hamiltonian,

$$H_{sp}(\mathbf{k}) = \begin{pmatrix} E_s^0 & 3t_s - \frac{3a^2 t_s}{4}k^2 & 0 & 0 & \frac{3ia\gamma_{sp}}{2}k_x & \frac{3ia\gamma_{sp}}{2}k_y \\ 3t_s - \frac{3a^2 t_s}{4}k^2 & E_s^0 & \frac{3ia\gamma_{sp}}{2}k_x & \frac{3ia\gamma_{sp}}{2}k_y & 0 & 0 \\ 0 & \frac{-3ia\gamma_{sp}}{2}k_x & E_p^0 & 0 & h(\alpha,\beta)^* & g(\beta)^* \\ 0 & \frac{-3ia\gamma_{sp}}{2}k_y & 0 & E_p^0 & g(\beta)^* & h(\alpha,-\beta)^* \\ \frac{-3ia\gamma_{sp}}{2}k_x & 0 & h(\alpha,\beta) & g(\beta) & E_p^0 & 0 \\ \frac{-3ia\gamma_{sp}}{2}k_y & 0 & g(\beta) & h(\alpha,-\beta) & 0 & E_p^0 \end{pmatrix}, \tag{S5}$$

$$h(\alpha,\beta) = \frac{3\alpha}{2} + \frac{3ia\beta}{4}k_y - \frac{3a^2\alpha}{8}k^2 - \frac{3a^2\beta}{16}(k_x^2 - k_y^2), \; g(\beta) = \frac{3ia}{4}k_y - \frac{3a^2\beta}{8}k_x k_y, \tag{S6}$$

where onsite energies and the nearest-neighbor hoppings are considered near the $\Gamma$ point. The spin degree of freedom is ignored. Here, $\alpha = t_\sigma + t_\pi$, $\beta = t_\sigma - t_\pi$, $a$ is the lattice constant, $E_s^0$ ($E_p^0$) is the on-site energy of the $s$ ($p$) orbitals, $t_s$ is the hopping integral of $s$ orbitals, $t_\sigma$ ($t_\pi$) is the $\sigma$ ($\pi$) hopping integral of $p$ orbitals, and $\gamma_{sp}$ is the $sp$ hybridization. When $\gamma_{sp} = 0$, $H_{sp}(\mathbf{k})$ becomes a block-diagonal matrix, of which $4 \times 4$ lower-right square matrix for $p_x$ and $p_y$ orbitals is decoupled from the $2 \times 2$ upper-left square matrix for $s$ orbitals. When $\gamma_{sp} \neq 0$, that is, when the $sp$ hybridization occurs, $H_{sp}(\mathbf{k})$ is not diagonal any more and all energy bands have both $s$ and $p$ characters. Nevertheless, when the onsite energies $E_s$ and $E_p$ for $s$ and $p$ orbitals are well separated compared to the $sp$ hybridization strength, there exist four energy bands with mostly $p$ characters. To derive an effective Hamiltonian for those $p$-character bands, we use the so-called Löwdin downfolding [57] technique.

The first step of this technique is to block-diagonalize $H_{sp}(\mathbf{k})$. For this purpose, we apply a unitary transformation $U = U_2 U_1$ where $U_1$ is a transformation to sublattice bonding and antibonding states and $U_2$ is a transformation that decouples $p$-character bands from $s$-character bands, which can be obtained by the Schrieffer–Wolff procedure. Then, $U H_{sp}(\mathbf{k}) U^\dagger$ is block-diagonalized into four blocks: antibonding $s$, bonding $s$, antibonding $p$, and bonding $p$. Finally, the effective Hamiltonian for $p$-character bands can be obtained by projecting $U H_{sp}(\mathbf{k}) U^\dagger$ onto a desired $p$-character block. For instance, the effective Hamiltonian $H_{p,-}(\mathbf{k})$ for the antibonding $p$-character block can be obtained by $H_{p,-}(\mathbf{k}) = P_p^- U H U^\dagger P_p^-$ where $P_p^-$ is the corresponding projection operator. After some algebra, one obtains

$$H_{p,-}(\mathbf{k}) = E_p I + \frac{\hbar^2 k^2}{2m^*} I + \frac{\eta \hbar^2}{2m^*} \begin{pmatrix} k_x^2 - k_y^2 & 2k_x k_y \\ 2k_x k_y & k_y^2 - k_x^2 \end{pmatrix}, \tag{S7}$$

where the matrix is represented in the transformed $(p_x, p_y)$ basis by $U$. Here, $E_p(E_s)$ is the $\Gamma$-point energy of the antibonding $p(s)$-character bands, $m^* = (8\hbar^2/3a^2)[\alpha + 4t_\sigma t_\pi/\alpha + 6\gamma_{sp}^2/(E_p - E_s)]^{-1}$ and $\eta = 3m^* a^2 \beta/8\hbar^2 + 9m^* a^2 \gamma_{sp}^2/4\hbar^2(E_p - E_s)$.

Here we keep terms up to the first order in $1/(E_p - E_s)$. Equation (S7) is the same Hamiltonian as $H(\mathbf{k})$ [Eq. (4)] in the main text. Note that $\eta$ has two contributions; one that arises from the difference between $\sigma$ and $\pi$ hopping integrals and the other that arises from the $sp$ hybridization. The effective Hamiltonian $H_{p,+}(\mathbf{k})$ for the bonding $p$-character block has similar features.

A physical system where the above analysis may apply is the $p$-doped graphane [31,32]. To be precise, there is a subtle issue since the mirror symmetry $z \to -z$ utilized in the above analysis is weakly violated in the $p$-doped graphane due to its buckling. According to the first-principles calculation [32], however, the two topmost valence bands of the $p$-doped graphane have mainly $p_x$ and $p_y$ characters and the buckling does not induce any qualitative change for the two topmost valence bands. Thus the above analysis successfully captures the main physics of $p$-orbital dynamics in the two topmost valence bands.

Besides the effective Hamiltonian for $p$-character bands, which differs from the "bare" Hamiltonian for $p$ orbitals due to the $sp$ hybridization, the effective operators for physical observables may differ from their "bare" versions due to the basis change during the derivation of the effective Hamiltonian through $U$. For an operator $A$, its $U$-induced correction can be calculated by $\Delta A_{p^-;U} = P_p^- U A U^\dagger P_p^- - P_p^- A P_p^-$. After some algebra, we verify that the $U$-induced corrections to the orbital angular momentum (OAM) and orbital angular position (OAP) operators are at least second order in $1/(E_s - E_p)$, which are neglected throughout our consideration. Therefore, in the large $E_s - E_p$ regime, these corrections can be safely discarded.

### III. DETAILS FOR THE ORBITAL DRIFT-DIFFUSION EQUATIONS

#### A. Linear response coefficients

We adopt $\tau_{\rm dp}^{-1} + \tau_m^{-1} \approx \tau_m^{-1}$, $\tau_{\rm of}^{-1} + \tau_m^{-1} \approx \tau_m^{-1}$ approximation for simplicity. However, qualitative features do not change. We calculate currents up to $\tau_m^{-1}$ order. Below, we list the nonzero components of the conductivity and diffusivity tensors and their physical meanings are discussed in the main text.

Nonzero components of the conductivity tensors are,

$$\chi_{xx}^0 = \chi_{yy}^0 = -\frac{e^2}{2\pi}\left(\frac{k_F^2 \tau_m}{m^*} + \frac{m^*}{\hbar^2 k_F^2 \tau_m}\right), \quad \chi_{xx}^3 = \chi_{yy}^3 = 0,$$

$$\chi_{yx}^1 = -\chi_{xy}^1 = -\frac{e^2}{4\pi}\left[\frac{(k_t^2 - k_r^2)\tau_m}{m^*} - \frac{2m^*}{\eta \hbar^2 k_F^2 \tau_m}\right], \quad \chi_{yx}^2 = -\chi_{xy}^2 = -\frac{e^2}{h}\frac{\ln\frac{k_t}{k_r}}{\eta}, \quad \text{(S8)}$$

where $k_F = \sqrt{2m^* E_F}/\hbar$, $k_r = \sqrt{2m_r^* E_F}/\hbar$ and $k_t = \sqrt{2m_t^* E_F}/\hbar$. We note that the two orbital bands contribute oppositely to the orbital-polarized conductivity ($\chi_{xx}^3, \chi_{yy}^3$), resulting in zero. However, this is an artifact of our simplified model where the scattering time for each band are the same. We have checked that if the scattering times are different for each band or if the Fermi surface is not strictly rotationally symmetric, $\chi_{ii}^3$ (and the coefficients that are presented to be zero below) can be nonzero.

Non-zero components of diffusivity tensors are below. For the charge chemical potential gradient ($j = 0$),

$$\xi_{xx}^{00} = \xi_{yy}^{00} = -\frac{e}{2\pi}\left(\frac{k_F^2 \tau_m}{m^*} + \frac{m^*}{\hbar^2 k_F^2}\frac{1}{\tau_m}\right), \quad \xi_{xx}^{30} = \xi_{yy}^{30} = 0,$$

$$\xi_{yx}^{10} = -\xi_{xy}^{10} = -\frac{e}{4\pi}\left[\frac{(k_t^2 - k_r^2)\tau_m}{m^*} - \frac{2m^*}{\eta \hbar^2 k_F^2 \tau_m}\right], \quad \xi_{yx}^{20} = -\xi_{xy}^{20} = -\frac{e}{h}\frac{1}{1-\eta^2}. \quad \text{(S9a)}$$

For the orbital torsion chemical potential gradient ($j = 1$),

$$\xi_{yx}^{01} = -\xi_{xy}^{01} = -\frac{e}{4\pi}\left[\frac{(k_t^2 - k_r^2)\tau_m}{m^*} - \frac{2m^*}{\eta \hbar^2 k_F^2 \tau_m}\right], \quad \xi_{yx}^{31} = -\xi_{xy}^{31} = -\frac{e}{4\pi}\left[\frac{\eta \tau_m (k_t^2 - k_r^2)}{m^*}\right]$$

$$\xi_{xx}^{11} = \xi_{yy}^{11} = -\frac{e}{4\pi}\left[\frac{\eta(k_t^2 - k_r^2)\tau_m}{m^*} + \frac{2m^*}{\eta^2 \hbar^2 k_F^2 \tau_m}\right], \quad \xi_{xx}^{21} = \xi_{yy}^{21} = -\frac{e}{h}\frac{\ln\frac{k_t}{k_r}}{\eta^2}. \quad \text{(S9b)}$$

For the OAM chemical potential gradient ($j = 2$),

$$\xi_{yx}^{02} = -\xi_{xy}^{02} = \frac{e}{h}\frac{\ln\frac{k_t}{k_r}}{\eta}, \quad \xi_{xx}^{12} = \xi_{yy}^{12} = \frac{e}{h}\frac{\ln\frac{k_t}{k_r}}{\eta^2}, \quad \xi_{xx}^{22} = \xi_{yy}^{22} = \frac{e}{2\pi}\frac{m^*}{\eta^2 \hbar^2 k_F^2 \tau_m}, \quad \xi_{yx}^{32} = -\xi_{xy}^{32} = 0. \quad \text{(S9c)}$$

For the orbital-state polarization chemical potential gradient ($j = 3$),

$$\xi_{yx}^{13} = -\xi_{xy}^{13} = \frac{e}{4\pi}\frac{\eta \tau_m (k_t^2 - k_r^2)}{m^*}, \quad \xi_{xx}^{33} = \xi_{yy}^{33} = -\frac{e}{2\pi}\frac{k_F^2 \tau_m}{m^*}, \quad \xi_{yx}^{23} = -\xi_{xy}^{23} = \frac{e}{h}\frac{\eta}{1-\eta^2}, \quad \xi_{xx}^{03} = \xi_{yy}^{03} = 0. \quad \text{(S9d)}$$



\begin{thebibliography}{99}
\bibitem{note1} Although the number of independent OAP operators are six, $L_x^2+L_y^2+L_z^2=2\hbar^2$ corresponds to the charge degree of freedom. Therefore, only five more operators are needed for complete description.
\bibitem{note2} Although $\{\sigma_i'\}$ form a set of Pauli matrices as spin, they have a different TR property. Under TR, $(\sigma_0'=I,\sigma_1',\sigma_2',\sigma_3')$ transform to $(\sigma_0',\sigma_1',-\sigma_2',\sigma_3')$ whereas the spin operators $(S_0=I,S_1,S_2,S_3)$ do to $(S_0,-S_1,-S_2,-S_3)$.
\bibitem{note3} Depending on situations, it may be preferable to define the basis as the one in the "lab frame", for example, $|p_x\rangle,|p_y\rangle$ instead of $|p_r\rangle,|p_t\rangle$. But in this Letter, we take the eigenstate basis since it is relevant to most transport measurements.
\bibitem{note4} Considering the time resolution of the TR-MOKE, the desired scale of the energy difference is on the order of 10 meV, which can be realized in a so-called zero-gap semiconductor or strained graphane where the energy difference is tunable.
\end{thebibliography}
\end{document}